\documentstyle[aps,epsfig,twocolumn,floats]{revtex}

\newcommand{\up}{\uparrow}
\newcommand{\down}{\downarrow}
\newcommand{\rhotilde}{\varrho}
\begin{document}
\draft   
\title{The interplay between shell effects and electron correlations in
quantum dots}
\author{Jens Harting, Oliver M{\"u}lken and Peter Borrmann}
\address{Department of Physics, Carl von Ossietzky University
Oldenburg, D-26111 Oldenburg, Germany}
\date{\today}
\maketitle
\begin{abstract}
We use the Path Integral Monte Carlo method to investigate the interplay
between shell effects and electron correlations in single quantum dots with up
to 12 electrons. By use of an energy estimator based on the hypervirial
theorem of Hirschfelder we study the energy contributions of
different interaction terms in detail.  We discuss under which conditions
the total spin of the electrons is given by Hund's rule, and the temperature
dependence of the crystallization effects.
\end{abstract}
\section{Introduction}
The advances in nanofabrication of the last years opened the goal to
build 2D quantum dots (QDs) and quantum dot molecules (QDMs)
--~artificial mesoscopic semiconductor structures of selectable shape
and size~-- as {\sl containers} for a controllable fixed number of
electrons \cite{Zhitenev:1997,Tarucha:1996}. Recently, depending on
the strength and shape of the effective confining potential, the
formation of spin density waves (SDWs) \cite{YL99,Koskinen:1997}
and Wigner crystals \cite{YL99,Egger:1999a} in QDs and QDMs
has been predicted by different groups with different theoretical
approaches. Hirose and Wingreen \cite{Hirose:1999} argue that SDWs
are reproducible artefacts of spin density functional calculations.
For a 2D parabolic confining potential the accordance of the
spin-configuration with Hund's Rule has been predicted by Koskinen,
Manninen, and Reimann \cite{Koskinen:1997} and questioned by
Yannouleas and Landman\cite{YL99}. All these effects are governed
by the intriguing interplay between shell effects, the pure coulomb
repulsion, and the fermionic repulsion due to the Pauli exclusion
principle and depend strongly on the values of the interaction
parameters in the commonly for single QDs assumed Hamiltonian
\begin{equation} \label{hamilton}
H = \sum_{i=1}^{N} \left( \frac{{\bf p}_i^2}{2 m^*}  
+ \frac{m^{*} \omega_0^2}{2} {\bf x}_i^2 \right)
+ \sum_{i<j=1}^N \frac{e^2}{\kappa | {\bf x}_i -{\bf x}_j|} 
\end{equation}
where $\kappa$ is the dielectric constant, $m^*$ is the effective
mass, and $\omega_{0}$ defines the strength of the confining
potential. 

Apart from the interesting physical questions that arise for quantum dots the
reliable prediction of their properties is an ultimate test of modern methods
in quantum chemistry. Due to the compared to atoms very shallow confining
potential long range electron interactions and correlations play an important
role in QDs and QDMs. Therefore it is misleading to name them artificial
atoms and molecules. Well established and very elaborate methods of
quantum chemistry might fail in describing them properly. Hartree-Fock and
spin density functional methods use single Slater determinants or sums of them
to approximate the many-body wave function. In spin-density functional methods
the approximation of the functional for the exchange correlation energy
\cite{Koskinen:1997,Pfannkuche:1993,Pino:1998} adds another source of
uncertainty and systematic errors to this approach.  The Path Integral Monte
Carlo  method (PIMC) used in this paper samples the full many-body wave
function instead.  

In contrast to density functional methods with PIMC it is possible to study the
temperature dependent properties of QDs. The reason why PIMC is not yet a
standard method of quantum chemistry is its numerical limitation due to the
fermion sign problem. The rapidly increasing power of modern computers resizes
this limitation. In sec.~\ref{Method} we briefly summarize our implementation of
PIMC and comment on how to limit the numerical deficiencies due to the fermion
sign problem.

We apply PIMC to calculate the electron density and two-particle
correlation functions for quantum dots with up to 12 electrons. To
compare to various experimental as well as to other theoretical studies we
use different dielectric constants $\kappa$ and strengths of the confining
parabolic potentials. The calculated addition energies are in very good
agreement with the experimental findings of Tarucha
et~al.~\cite{Tarucha:1996}.

For $N=6$ we investigate the temperature
dependence of the Wigner crystallization~(WC).
 
\section{Numerical method} \label{Method}
For a system of $N$ electrons with
position eigenket ${\mid\vec{x}_i,s_i\rangle}$ ($s_i =\pm
\frac{1}{2}$ for spin-up and spin-down electrons) in an external
potential the Feynman path integral can be written as  
\cite{takahashi84a,takahashi84b,bh93}
\begin{eqnarray} \label{pfad}
Z &=& \int 
\left[
\prod_{\gamma=1}^{M} \prod_{i=1}^{N} {\rm d}{\vec x}_i(\gamma)
\right]
\prod_{\delta=1}^{M} \det(A(\delta,\delta+1))\\
&\times&\exp\left(-\frac{\beta}{M} \sum_{\alpha=1}^{M}
V(\vec{x}_1(\alpha),\ldots,\vec{x}_N(\alpha))\right) + 
{\cal O}\left(\frac{\beta^3}{M^2}\right)\nonumber
\end{eqnarray}
with  
\begin{eqnarray}
&&(A(\alpha,\alpha+1))_{i,j} \\
&&=\left\{ 
\begin{array} {l@{\quad:\quad}r}
\langle \vec{x}_{i}(\alpha)\mid 
\exp\left( -\frac{\beta}{M} \frac{{\bf p}^2}{2m} \right) \mid \vec{x}_{j}(\alpha+1)
\rangle & s_i = s_j\\
0       & s_j \neq s_j
\end{array}
\right.\nonumber
\end{eqnarray}
and the boundary condition $\vec{x}_{j}(M+1)=\vec{x}_{j}(1)$.
$M$ is the number of so-called {\sl timeslices} of the Feynman paths. 
In the limit $M\rightarrow \infty$ Eq.~(\ref{pfad}) becomes exact.
For quantum dots the space dimension is d=2 and  
the (2NM)-dimensional integral given in (\ref{pfad}) can be evaluated 
by standard Metropolis Monte Carlo techniques. Due to the determinant 
the integrand is not always positive and the expectation value of 
an observable $X(\bf x)$ depending only on position operators has to be
calculated using  
\begin{equation} \label{sign}
\langle X \rangle = \frac{\sum_{g=1}^G X_g {\rm sign}(W_g)}{\sum_{g=1}^G
{\rm sign}(W_g)}
\end{equation}  
where $X_g$ is the value of the observable $X$ and $W_g$ is the value 
of the integrand in (\ref{pfad}) in the g'th Monte Carlo step.
Eq. (\ref{sign}) reveals a severe problem connected with the path
integral for fermions which is commonly denoted as the {\sl fermion
sign problem}  (see e.g. \cite{cephe3,lyuba,Morgenstern}). 
It can be shown that the ratio between integrands with positive sign
($W^+$) and negative sign ($W^-$) is approximately given
by~\cite{Morgenstern,newman}
\begin{equation}
\frac{W^+-W^-}{W^+ + W^-} \sim \exp(-\beta (E_F-E_B) )
\end{equation}
where $E_F$ and $E_B$  are the ground state energies of the Fermi system
and the corresponding Bose system. It is now obvious that the
statistical error in (\ref{sign}) grows rapidly for small temperatures
$T$. Moreover the energy difference $(E_F-E_B)$ will grow with 
increasing system size causing an increase of the statistical error.

Within PIMC the calculation of the kinetic energy expectation 
value is another critical task. This is merely due to the fact that 
the Monte Carlo calculation is usually done in position space and that 
the discretization of the paths allows a number of different approaches
to calculate the expectation value of a momentum dependent operator.
A number of various different energy estimators has been discussed 
in the past \cite{cb89,gj88,ktl88}. 

To avoid these difficulties we developed a procedure which allows the 
calculation of all energy expectation values from the knowledge
of the pair correlation functions 
\begin{equation}
\Gamma_{i,j}(r)=\langle\delta(r-\mid \vec{x}_i-\vec{x}_j\mid)\rangle
\end{equation}
and the radial density functions per electron
\begin{equation}
\rho_{i}(r) = \frac{1}{2 \pi r} \langle \delta( r - \mid \vec{x}_i\mid)
\rangle = \frac{1}{2\pi r} \rhotilde(r) \;,  
\end{equation}
where $\rhotilde$ is the probability of finding electron $i$ in distance $r$
from the center.

Due to the particle symmetry we have 
\begin{equation}
\Gamma_{i,j}(r) = 
\left\{ 
\begin{array}{r@{\quad:\quad}l}
\Gamma^{\up\up}(r) & s_i=s_j=+\frac{1}{2}\\
\Gamma^{\down\down}(r) & s_i=s_j=-\frac{1}{2}\\
\Gamma^{\up\down}(r) & s_i\neq s_j
\end{array}
\right.
\end{equation}
and 
\begin{equation}
\rho_i(r) = 
\left\{ 
\begin{array}{r@{\quad:\quad}l}
\rho^{\up}(r) &  s_i=+\frac{1}{2}\\
\rho^{\down}(r) &  s_i=-\frac{1}{2}
\end{array}
\right..
\end{equation}
Utilizing the hypervirial theorem of Hirschfelder \cite{hirschfelder}
the energy can be written as a sum of ten parts~\cite{hbsh96} 
\begin{eqnarray} \label{energy}
E &=& E^{\up}_{{\rm kin}} + E^{\down}_{{\rm kin}} +E^{\up\up}_{{\rm kin}}
+ E^{\down\down}_{{\rm kin}} + E^{\up \down}_{{\rm kin}}\\
&&+ E^{\up}_{{\rm pot}} + E^{\down}_{{\rm pot}} +E^{\up\up}_{{\rm pot}} 
+ E^{\down\down}_{{\rm pot}} + E^{\up \down}_{{\rm pot}}\nonumber \\
&=& 
\frac{N^{\up}}{2} \int_{0}^{\infty} {\rm d}r\,\rhotilde^{\up}(r) r \partial_r V_1(r)
+\frac{N^{\down}}{2} \int_{0}^{\infty} {\rm d}r\, \rhotilde^{\down}(r) r \partial_r V_1(r)
\nonumber \\
&&+ \frac{N^{\up}(N^{\up}-1)}{4} \int_{0}^{\infty} {\rm d}r\,
\Gamma^{\up\up} (r)\, r \, \frac{\partial V_2(r)}{\partial r}\nonumber\\
&&+ \frac{N^{\down}(N^{\down}-1)}{4} \int_{0}^{\infty} {\rm d}r\,
\Gamma^{\down\down} (r)\, r \, \frac{\partial V_2(r)}{\partial r}
\nonumber \\
&&+ \frac{N^{\down} N^{\up}}{2} \int_{0}^{\infty} {\rm d}r\,
\Gamma^{\up\down} (r)\, r \, \frac{\partial V_2(r)}{\partial r} \nonumber \\
&& 
+\frac{N^{\up}}{2} \int_{0}^{\infty} {\rm d}r\, \rhotilde^{\up}(r)  V_1(r)
+\frac{N^{\down}}{2} \int_{0}^{\infty} {\rm d}r\, \rhotilde^{\down}(r) V_1(r)
\nonumber \\
&&+\frac{N^{\up}(N^{\up}-1)}{2} \int  {\rm d}r\, \Gamma^{\up\up}(r)
V_2(r)\nonumber\\
&&+\frac{N^{\down}(N^{\down}-1)}{2} \int  {\rm d}r\, \Gamma^{\down\down}(r) V_2(r)
\nonumber \\
&&+N^{\up} N^{\down} \int  {\rm d}r\, \Gamma^{\up\down}(r) V_2(r)
\nonumber
\end{eqnarray}
While in density functional approaches the calculation of the kinetic energy
and the exchange correlation energy is a major topic and subject to permanent
discussion, within the path integral approach these energies are included in
a natural way.\\
However, the systematic error arising from the limited number of timeslices $M$
and the statistical error of the Monte Carlo calculation have to be controlled
carefully.  We checked our algorithm extensively using eight non-interacting
fermions in a parabolic trap as a test system. We found that at low
temperatures where the ratio of signs is around 0.99, convergence can only be
achieved obeying the following rules: 1) The determinants have to be calculated
very accurately using a more costly algorithm with pivoting. 2) The
completely uncorrelated generation of
the Monte Carlo steps is essential, i.e. the coordinate to
be moved should be chosen randomly. Moving the particle coordinates
using always the same sequence produces inaccurate results. 3) A good random
number generator with a {\sl completely} uncorrelated sequence in all
significant bits of a 64 bit real number should be applied. We therefore
developed a 53 bit random number of Marsaglia-Zaman type\cite{Borrmann:2000}
instead of using one of the standard 24 or 32 bit random number generators coming
with standard system libraries. 4) Further, to improve the convergence a
number of different Monte Carlo steps can be applied, i.e. moving single
time slices, moving complete particle paths and parts of a path.\\
Our Fortran-code is completely parallelized using MPI and Lapack. 
\section{Results} 
To compare our PIMC calculations to experimental data  we calculated 
the addition energies
\begin{equation}
\Delta E = E_{N+1} - 2 E_N + E_{N-1}
\end{equation} 
of a QD with up to 11 electrons using the material constants $m^{*} = 0.067m$
and $\kappa$ = 12.9 for GaAs  as given by Hirose and Wingreen
\cite{Hirose:1999}. It is assumed that these parameters mimic the experimental
setup of Tarucha et~al.~\cite{Tarucha:1996} reasonably well. The strength of
the harmonic potential is fixed at $\hbar \omega_0$ = 3.0 meV. The resulting
effective atomic units are $E_H^*$ = 10.955 meV for the Hartree energy and
$a_0^*$ = 10.1886 nm for the Bohr radius. The Boltzmann constant is $k_B =
7.8661\times10^{-3} E_H^*/{\rm K}$.

\begin{figure}[ht]
\centerline{\epsfig{file=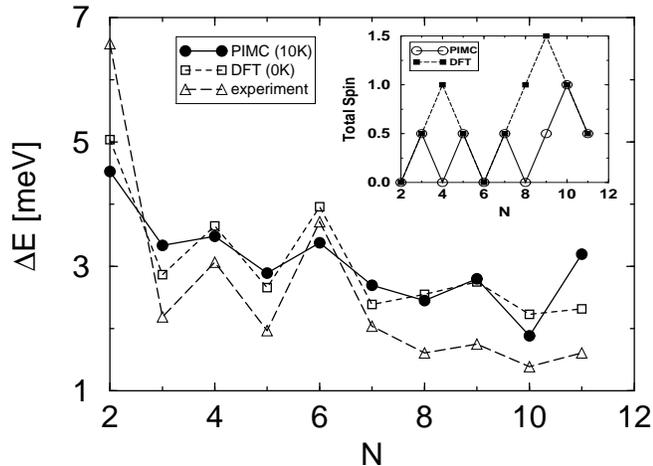,height=6.5cm}}
\caption{Addition energies for quantum dots with up to 11 electrons. 
The circles indicate the results
from our path integral calculations at 10 K, the squares are results of 
spin density functional calculations of Hirose and
Wingreen, and the triangles are the experimental results 
of Tarucha et~al. The error bars for PIMC would be of the size of the
solid circles and are therefore omitted. 
}
\label{addfig}
\end{figure}

We performed PIMC simulations for quantum dots  with different spin
configurations at a fixed temperature of 10 K. Due to the fermion sign problem
the number of Monte Carlo steps necessary to push the statistical error of the
total energy, which has been calculated properly from 25 uncorrelated 
subsequences of MC steps, into the range of 0.1 percent is extremely high.
The number of Monte Carlo steps ranged between 2.5 billion steps per particle 
coordinate for
$N\le6$ and about 10 billion steps for $N=12$.  Fig.~\ref{addfig} displays the
addition energies for quantum dots with up to 11 electrons. The circles
indicate the results from our path integral calculations at 10 K, the squares
are results of spin density functional calculations of Hirose and
Wingreen~\cite{Hirose:1999}, and the triangles are the experimental results of
Tarucha et~al.~\cite{Tarucha:1996}. 
Both theoretical calculations reproduce the general $N$-dependence of
the addition energies in great detail. Tarucha et~al. give an estimate
of the electron temperature in their experiments of $T=0.2$~K.
For computational reasons our PIMC calculations are performed at 10 K
and it cannot be expected that the absolute energy values agree as
well as the 0~K DFT calculations with the experimental results.
However, it should be noted that PIMC correctly  predicts the drop in the
addition energy from $N=7$ to $N=8$ while the DFT calculations fail
at this point. 

The inset in Fig.~\ref{addfig} displays the total spins of the spin
configurations with lowest energy as found in DFT and PIMC at 10 K. In
DFT calculations (0 K) the spin configuration of the ground state is determined
by Hund's rule for up to 22 electrons. In contrast, in our PIMC calculation
at 10 K the total spin is not always in accordance with Hund's rule.  
For  $N=4$ we checked the temperature dependence of the spin configuration.  At
5 K the energy of the spin 0 configuration is 0.01 $E_{H}^*$ higher than the
spin 1 energy indicating a temperature dependence of the favored spin
configuration.

As an important fact we note that the $N$-dependence of the addition energies
is not affected by the actual spin configuration. The situation is quite
similar to that  in transition metal clusters with extreme small energy
differences between states with significantly different magnetic moments \cite{Lee:1993}.

\begin{figure}[ht]
\centerline{\epsfig{file=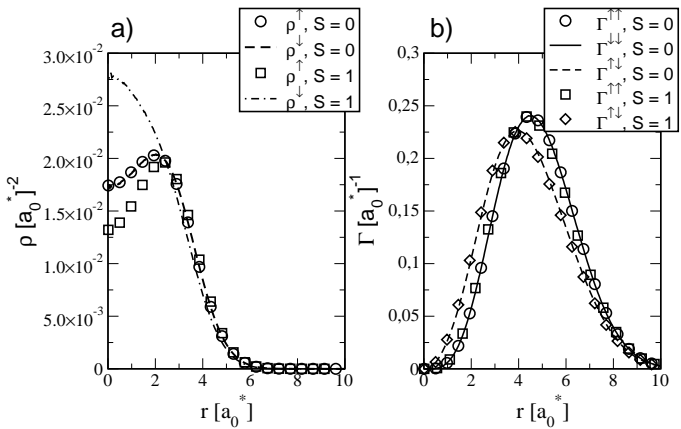,height=7cm}}
\caption{Radial density per electron (a) and pair correlation functions 
(b) for 4 electrons and total spins $S$=0 and $S$=1. 
The material constants are $\kappa$ = 12.9 and $\hbar \omega$ = 3.0 meV. }
\label{rhoGamma}
\end{figure}

As can be inferred from Fig.~\ref{rhoGamma}(a) the radial spin densities
are significantly different for both spin configurations. The total
potential energy for the spin 1 configuration is about 0.07 meV lower
than that of the spin~0 configuration. At 10 K this is overcompensated
by an 0.27 meV higher kinetic energy (see Tab.~\ref{Etable}).  Although
the kinetic and potential energies for different total spins
significantly differ the total energies are almost equal. 
Similar situations are found for larger $N$. 

For convenience and easy comparision  we determined the value of the
dimensionless density parameter $r_s$, which is sometimes used 
to characterize quantum dots (see e.g. \cite{Egger:1999a}) to be 
$r_s$=4.19 for $N=4$.  

\begin{table}[ht]
\begin{tabular}{|l|c|c|c|c|c|c|}
\hline
&$N^\uparrow$=2,$N^\downarrow$=2&
$N^\uparrow$=3,$N^\downarrow$=1&
$N^\uparrow$=5,$N^\downarrow$=4&
$N^\uparrow$=6,$N^\downarrow$=3\\
\hline\hline
$E_{\rm tot}$&40.83&41.03&169.25&169.82\\
$E_{\rm kin}$&7.33&7.60&18.97&19.57\\
$E_{\rm pot}$&33.50&33.43&150.28&150.26\\
$E_{\rm W}$&0.18&0.19&0.11&0.12\\
\hline
$E_{\rm kin}^\uparrow$&8.03 (8.03)&12.66 (8.44)&35.44 (14.18)&43.74 (14.58)\\
$E_{\rm kin}^\downarrow$&8.03 (8.03)&3.55 (7.10)&27.30 (13.65)&19.39 (12.93)\\
$E_{\rm kin}^{\uparrow\uparrow}$&-1.31 (-2.62)&-3.97 (-2.65)&-11.20
(-2.24)&-16.87 (-2.25)\\
$E_{\rm kin}^{\uparrow\downarrow}$&-6.11 (-3.05)&-4.64 (-3.09)&-25.84
(-2.58)&-23.31 (-2.59)\\
$E_{\rm kin}^{\downarrow\downarrow}$&-1.31 (-2.62)& 0.00 (0.00)&-6.72
(-2.24)&-3.39 (-2.26)\\
\hline
$E_{\rm pot}$&33.50&33.43&150.28&150.26\\
$E_{\rm pot}^\uparrow$&8.03 (4.01)&12.66 (4.22)&35.44 (7.09)&43.74 (7.30)\\
$E_{\rm pot}^\downarrow$&8.03 (4.01)&3.55 (3.55)&27.30 (6.82)&19.39 (6.46)\\
$E_{\rm pot}^{\uparrow\uparrow}$&2.62 (2.62)&7.94 (2.65)&22.40
(2.24)&33.73 (2.25)\\
$E_{\rm pot}^{\uparrow\downarrow}$&12.21 (3.05)&9.28 (3.09)&51.69
(2.58)&46.62 (2.59)\\
$E_{\rm pot}^{\downarrow\downarrow}$&2.62 (2.62)&0.00 (0.00)&13.44
(2.24)&6.78 (2.26)\\
\hline
\end{tabular}
\caption{Kinetic and potential energies as well as
$E_{\rm W}$=$E_{\rm kin}/E_{\rm tot}$ in meV for different electron
configurations at 10 K (the numbers in brackets are the single particle energies).}
\label{Etable}
\end{table}

The energies given in Tab.~\ref{Etable} correspond to the integrals in
Eq.~\ref{energy}. The total kinetic energy which is the sum of all 
$E_{\rm kin}^x$ terms is always positive while some of the addends might be 
negative. Tab.~\ref{Etable} reveals that larger total spins result in 
larger kinetic energies. The total potential energy is almost unchanged, 
the larger contribution from the trap potential is compensated by a smaller
contribution from the Coulomb repulsion. We note that the ratio $E_{\rm W}$
between the kinetic energy and the total energy is considerably
larger for $N=4$ than for $N=9$ reflecting the looser binding of the smaller
system.   

Next we consider the dependence of the Wigner crystallization on the
temperature and the choice of the material constants.  The localization of the
electrons in space is commonly referred as Wigner crystallization. For quantum
dots the occurrence of well separated humps in the radial electron density and
the pair correlation functions has been interpreted as WC. However, it is a
nontrivial task to find a general parameter identifying if an electron system
is crystallized or not. From a solid state physics point of view the electrons
should have a low mobility, i.e. a small kinetic energy, and should not
interchange their lattice positions. For fermions the localization of single
electrons does not make any sense, and, as stated above, even the decomposition
of the many body wave function in sums of determinants of single particle wave
functions is  probably a too rough approximation for QDs. These facts limit
the analogies between crystallization in solids and electron systems and make
the term {\sl crystallization } itself somehow misleading. We therefore view
Wigner crystals as states of the many body wave function with a relatively low
kinetic energy.

\begin{figure}[ht]
\centerline{\epsfig{file=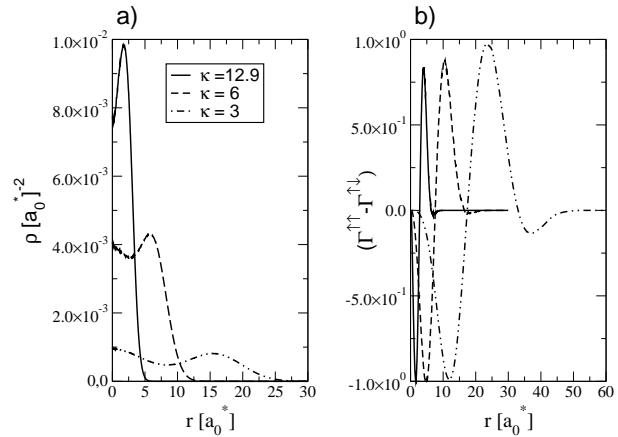,height=6.5cm}}
\caption{Radial density per electron (left) and pair correlation 
functions (right) for $N=6$, $S=0$, $\hbar\omega_0$=5~meV and dielectric
constants  $\kappa$=3.0, 6.0, and 12.9 at 10~K. The radial density for
$\kappa$=12.9 is scaled by a factor of 3 and the pair correlation functions
are scaled to have a maximum value of $\pm$1.}
\label{Wignerkappa}
\end{figure}

First we consider the strength of the Wigner crystallization depending
on the choice of the interaction parameters. Fig. \ref{Wignerkappa}
displays the radial densities and pair correlation functions for six
electrons with $S$=0, $\hbar \omega$=5~meV and $\kappa$=3.0, 6.0, and
12.9. Of course, for stronger electron repulsions (small $\kappa$) the
electron distributions are broadened. The qualitative picture of the
distributions is merely the same. For all $\kappa$ shell effects
indicated by off-center maximums of the radial density occur. However,
only for $\kappa$=3 and 6 we observe a maximum at $r$=0. From our point
of view it cannot definitely be decided from this figure if a system is
Wigner crystallized or not. As a parameter reflecting the strength of
the WC we employ the ratios between the kinetic and the total energies
$E_{\rm W} = E_{\rm kin}/E_{\rm tot}$  which are 0.07, 0.10 and 0.14 for
$\kappa$=3.0, 6.0 and 12.9.  Although the radial distribution function
for $\kappa=$12.9 is quite narrow, the relative mobility for the
electrons indicated by $E_{\rm W}$ is twice as large as for $\kappa=$3.
The underlying physical process can be understood intuitively. Due to the
stronger electron-electron repulsions the electrons are fixed in an
energetical favorable geometric configuration and as a consequence
thereof the relative kinetic energy is reduced and the difference between 
the pair correlation functions of equal and oppposite spin almost vanishes 
(see Fig.~\ref{Wignerkappa}).
\begin{figure}[ht]
\centerline{\epsfig{file=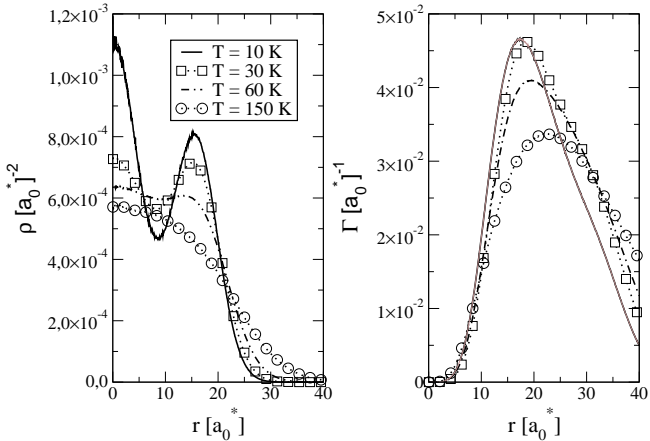,height=7.0cm}}
\caption{Radial density (left) and pair correlation functions (right) for
$N=6$, $S=0$, $\hbar\omega_0$=5~meV, $\kappa=3.0$, and temperatures $T$ =
10, 30, 60, and 150~K.}
\label{WignerrG}
\end{figure}
It is an interesting and to our knowledge open question, if the crystallization
of electrons can be viewed as a phase transition.  We therefore consider next
the temperature dependent properties of a quantum dot with N=6, S=0, $\kappa$=3
and $\hbar \omega$=5~meV.  The results for temperatures between 10 and 150~K
are presented in Tab.~\ref{Ekintable}. Most notably $E_{\rm W}$ increases
relatively smoothly from 0.07 at 10 K to 0.22 at 150 K. Within our numerical
accuracy the caloric curve does not show any evidence of a phase transition.
The transition from a crystallized state to an electron {\sl fluid} seems to be
squashy. Of course, from our calculations we cannot exclude that a phase
transition exists for larger $N$ or different interaction parameters.
Fig.~\ref{WignerrG} displays the radial electron densities and the total
electron pair correlation function for different temperatures. Up to 60 K the
radial density shows clear geometric structure effects with two maximums while
at 150 K only a smooth curve resembling to a simple gaussian remains.
\begin{table}[ht]
\begin{tabular}{|l|c|c|c|c|}
N$^\uparrow$=3, N$^\downarrow$=3& $E_{\rm W}$ &$E_{\rm kin}$ [meV]&$E_{\rm pot}$
[meV]&$E_{\rm tot}$ [meV]\\
\hline\hline
$T$ = 10 K&0.07&17.70&246.53&264.32\\
$T$ = 30 K&0.08&22.93&249.53&272.46\\
$T$ = 60 K&0.12&35.39&257.67&293.06\\
$T$ = 90 K&0.16&49.44&267.01&316.46\\
$T$ = 120 K&0.19&64.10&276.84&340.93\\
$T$ = 150 K&0.22&78.49&286.88&365.35\\
\end{tabular}
\caption{Kinetic and potential energies for different temperatures and spin configuration
$N^\uparrow$=3, $N^\downarrow$=3. The Hartree energy is $E_H^*$ = 202.558 meV,
$\kappa$ = 3 and $\hbar\omega_0$ = 5.0 meV. $E_{\rm W}$ is the ratio between the 
kinetic and the total energy.}
\label{Ekintable}
\end{table}
\section{Conclusion}
In conclusion, we have found that despite of the notorious fermion sign problem 
PIMC is capable to answer interesting questions for strongly correlated 
electron systems like QDs and QDMs. For QDs PIMC reproduces correctly 
the experimental addition energies. Our temperature dependent calculations 
give new insights into the process of WC. For the 2-dimensional QDs a ratio
$E_{\rm W} = E_{\rm kin}/ E_{\rm tot}$ below 0.1 seems to indicate WC both 
for $\kappa$ and temperature dependent calculations. However, reagarding this 
aspect a more firm classification parameter e.g. similary to the Lindemann 
criterion is desirable.

A comparision to other QMC methods seems to be in order here.  Even our most
complicated simulations took less than two hours on a Cray T3E with 62
processors.  Taking the advantages of modern computer power and optimized
software, the {\sl brute force} PIMC applied here is able to produce very
precise results. Different algorithms have been published to improve the
performance of path integral methods. A recent one is the {\sl Multi-Level
Blocking} method published by Mak et~al.~\cite{Mak:1998}. Simulations using
this algorithm are expected to converge better than our direct treatment
since the fermion sign problem is avoided partially. The published results
of the
treatment of quantum dots by Egger et~al.~\cite{Egger:1999a} do not confirm
this expectation. Obviously the advantages of the new method are more than
compensated by other numerical problems, maybe due to the energy estimator used
\cite{Rpimc}. Nevertheless, a combination of the Multi-Level-Blocking method and our technique might result in a very powerful tool.
\section{Acknowledgments}
We wish to thank the {\sl Regionales Rechenzentrum Niedersachsen}
and the {\sl Konrad Zuse Institut Berlin} for their excellent computer 
support and E.~R.~Hilf for stimulating discussions.

\end{document}